\documentclass[12pt]{article}

\usepackage{graphicx}
\usepackage{hyperref}
\usepackage{mathrsfs}
\usepackage{latexsym}
\usepackage{amsfonts}
\usepackage{amsmath}
\usepackage{esint}
\usepackage{amsxtra}

\def \m{\mbox}
\def \be{\begin{equation}}
\def \ee{\end{equation}}
\def\beq{\begin{eqnarray}}
\def\eeq{\end{eqnarray}}
\def \ba{\begin{array}}
\def \ea{\end{array}}
\def\sin{\mbox{sin}}
\def\cos{\mbox{cos}}

\def\nn{\nonumber}

\def \f{\frac}

\def \p{\partial}

\begin{document}
\vspace*{-.6in}
\thispagestyle{empty}
\baselineskip = 18pt

\vspace{.5in}
\vspace{.5in}
{\LARGE
\begin{center}
Magnetic expansion of Nekrasov theory: \\the  SU(2) pure gauge
theory
\end{center}}

\vspace{1.0cm}

\begin{center}

Wei He,\footnote{weihe@nankai.edu.cn} \qquad Yan-Gang Miao\footnote{miaoyg@nankai.edu.cn} \\
\vspace{1.0cm}\emph{Department of Physics, Nankai University,
Tianjin 300071, China}
\end{center}
\vspace{1.0cm}

\begin{center}
\textbf{Abstract}
\end{center}
\begin{quotation}
\noindent It is recently claimed by Nekrasov and Shatashvili that
the $\mathcal {N}=2$ gauge theories in the $\Omega$ background with
$\epsilon_1=\hbar, \epsilon_2=0$ are related to the quantization of
certain algebraic integrable systems. We study the special case of
SU(2) pure gauge theory, the corresponding integrable model is the
A$_1$ Toda model, which reduces to the sine-Gordon quantum mechanics
problem. The quantum effects can be expressed as the WKB series
written analytically in terms of hypergeometric functions. We obtain
the magnetic and dyonic expansions of the Nekrasov theory by
studying the property of hypergeometric functions in the magnetic
and dyonic regions on the moduli space. We also discuss the relation
between the electric-magnetic
duality of gauge theory and the action-action duality of the integrable system.
\end{quotation}

\newpage

\pagenumbering{arabic}

\section{Introduction}

The nonperturbative properties of quantum field theories have been
one of the most active research subjects during the past few
decades, we have known a lot of information through various
analytical or numerical methods. The four dimensional Yang-Mills
theory stands as one of the few most attractive field models. One of
the milestones of studying the supersymmetric gauge theories is the
Seiberg-Witten solution of the four dimensional $\mathcal {N}=2$
gauge theory \cite{SW9407}, which results in a fully analytic
understanding of a large class of supersymmetric gauge theories.
Their solution is based on typical features of supersymmetric gauge
theories, i.e. the holomorphic structure of prepotential and the
electric-magnetic duality of gauge theory. By analyzing the vacuum
structure of the moduli space and the related monodromy problem,
Seiberg and Witten discovered that the low energy physics of the
gauge theory is encoded in a geometric object, an elliptic curve,
the prepotential can be obtained through the periods of a
holomorphic differential one form along the two conjugate homology
cycles. The  periods can be written as hypergeometric functions on
the moduli space, they manifest the electric-magnetic duality in a
very explicit way: the electric-magnetic duality group of the gauge
theory is the same as the discontinuous reparametrization group of
the elliptic curve. The solution is valid on the whole moduli space.
In some region the theory is a weakly coupled electric theory, in
some region the electric theory is strongly coupled, but it can be
reformulated as a weakly coupled magnetic theory. By choosing
suitable quantities as the fundamental degrees of freedom, we can
either expand the effective action in terms of the electric fields
or in terms of the magnetic (or dyonic) fields. Subsequent works
have extended the solution to the $\mathcal {N}=2$ theory with more
general gauge groups and with matters, it is also found that these
solution can be interpreted in the context of string theory, see
review \cite {A-GH9701, L9611}.

The original work of Seiberg and Witten is reinterpreted in
\cite{NN03} from a different viewpoint. The hard part of solving the
$\mathcal {N}=2$ gauge theory is the sum of the instanton
contributions, but the multi-instanton measure on moduli space grows
very complicated as the number of instantons increases, only the
first few multi-instanton contributions have been calculated
directly. The problem is solved through the localization technique,
this can be achieved only after embedding the $\mathcal {N}=2$ gauge
theory into the so called $\Omega$-background \cite{NN03, NO03,
LNS97, MNS97, LNS98, MNS98}. The $\Omega$-background is a twist of
the $\mathbb{R}^4$ bundle characterized by two complex parameters
$\epsilon_1,\epsilon_2$. The partition function of this theory can
be expressed in a compact form as a contour integral and can be
analytically performed to arbitrary order in the instanton
expansion. The instanton part of the Seiberg-Witten theory
$\mathcal{F}^{inst}(a,m,q)$ can be obtained through the Nekrasov
partition function by the limit $\epsilon_1=-\epsilon_2=\hbar\to0$.

The Nekrasov theory not only gives Seiberg-Witten theory a more
mathematically solid explanation, and the theory is also important
by itself, its rich structure is still largely unknown. One of the
still mysterious aspects of Nekrasov theory is its modular property,
i.e. the electric-magnetic duality property. In the original
Seiberg-Witten formulation of the solution, the electric-magnetic
duality is manifestly realized, the hypergeometric function is well
defined on the whole moduli space coordinated by $u$, and we can get
asymptotic expansion near $u=\infty$ and $\pm\Lambda^2$ which
correspond to electric region and magnetic (dyonic) region,
respectively. But the Nekrasov theory is defined in the electric
region, its partition function does not directly depend on the
moduli space coordinate $u$, and it is not clear how the
electric-magnetic duality works.  It is interesting to find a way to
study the magnetic (or dyonic) expansion of Nekrasov theory. Some
earlier works concern this problem appear in \cite{HK0605}, the
authors study the Nekrasov theory with
$\epsilon_1=-\epsilon_2=\lambda$ which is related to topological
string theory and matrix model \cite{KMT0206}. The general case of
$\epsilon_1, \epsilon_2$ remains unknown.

A rather interesting discovery is recently presented in
\cite{NS0908, NS0901}, where the relation between the Nekrasov
theory and the quantization of algebraic integrable system is
established. The correspondence between $\mathcal {N}=2$ gauge
theories and the classical integrable systems has been extensively
studied soon after Seiberg-Witten theory. It was noted in
\cite{GK3M95, DW9510} that the Seiberg-Witten solution of gauge
theory is related to the classical integrable system, more
precisely, the Seiberg-Witten curve of the gauge theory is identical
to the spectral curve of the classical integrable system if we
suitably identify physical quantities on the two sides, see review
\cite{HP99, M1999}. In \cite{NS0908}, the authors develop the
correspondence to the quantum level, they claim that the
$\Omega$-background twisted gauge theory with
$\epsilon_1=\epsilon,\epsilon_2=0$ is related to the quantization of
the corresponding classical integrable system, and the
$\epsilon$-parameter of the gauge theory is identified with the
Plank constant $\hbar$. Here the ``quantization'' refers to the
integrable system side, on the gauge theory side it corresponds to
higher order $\epsilon$ twist expansion. Some evidence supporting
the correspondence is presented in \cite{MM0910, WH1004}, they
consider the special case of SU(2) pure Yang-Mills theory which is
related to the A$_1$ Toda integrable system, the A$_1$ Toda system
reduces to the sine-Gordon quantum mechanics problem on the complex
plane. It is shown that the energy spectrum and wave function of the
quantized mechanical model give the results consistent with the
requirement of Nekrasov theory. Discussion on more general cases is
presented in \cite{MM0911}.

In \cite{MM0910}, the authors found that the higher order $\hbar$
corrections can be obtained via acting on certain higher order
differential operators on the leading order result, i.e. the
Seiberg-Witten solution. This fact indicates that the $\hbar$
corrections can be also expressed compactly in hypergeometric
functions which are valid on the whole moduli space. We use this
observation to study the magnetic expansion of Nekrasov theory with
$\epsilon_1=\hbar, \epsilon_2=0$, by expanding higher order contour
integrals in the magnetic region on the moduli space. Dyonic
expansion is also obtained, it manifests a similar pattern with the
magnetic case. In fact, the magnetic and dyonic expansions are
related by a $\mathbb{Z}_2$ symmetry, therefore, we mainly discuss
the electric-magnetic duality of the system.

\section{Higher order contour integrals}

In the Seiberg-Witten solution of $\mathcal {N}=2$ SU(2) pure gauge
theory the quantum moduli space is a complex quantity $u$ of mass
dimension two on which there are three singularities at $u=\infty,
\pm\Lambda^2$ which correspond to the electric region and magnetic
(dyonic) region, respectively. $\Lambda$ denotes the dynamical
generated scale of the gauge theory, for simplicity we set the scale
$\Lambda=1$ and it can be restored by dimensional analysis at last.
The moduli space is the quotient of the upper half plane $H$ by
$\Gamma(2)$, where $\Gamma(2)$ is subgroup of $SL(2, \mathbb{Z})$
congruent to 1 modulo 2. The low energy effective action is
described by the Seiberg-Witten curve whose moduli space is exactly
$H/\Gamma(2)$: \be y^2=(x^2-1)(x-u)\ee and the corresponding
Seiberg-Witten differential \be
d\lambda(u,x)=\f{\sqrt{2}}{2\pi}\f{\sqrt{x-u}}{\sqrt{x^2-1}}dx.\ee
Then $a^{(0)}$ and $a_D^{(0)}$ are integrals of $d\lambda$ along the
conjugate circles $\alpha$ and $\beta$, respectively, \be
a^{(0)}=\oint_\alpha d\lambda, \qquad a_D^{(0)}=\oint_\beta
d\lambda,\label{SWaad}\ee where $a^{(0)}$ is the vacuum expectation
value of the scalar field, and $a_D^{(0)}$ is the dual quantity. On
the moduli space, the two cycles $\alpha$ and $\beta$ correspond to
the integral contours encircling branch point pairs $(-1,+1)$ and
$(+1,u)$, respectively. The result can be written in terms of
hypergeometric
function, \beq a^{(0)}(u)&=&\sqrt{2(u+1)}F(-\f{1}{2},\f{1}{2},1;\f{2}{u+1})\nn\\
a_D^{(0)}(u)&=&\f{i}{2}(u-1)F(\f{1}{2},\f{1}{2},2;\f{1-u}{2}).\label{aaD0}
\eeq They are well defined on the whole moduli space.

At each singularity electric (or magnetic/dyonic) particles become
massless, if we treat the corresponding massless particles as
fundamental degrees of freedoms, the theory is weakly coupled in the
region near the singularity. Near $u=\infty$, the theory is weakly
coupled electric theory, and the low energy effective prepotential
$\mathcal{F}^{(0)}$  is obtained from \be a_D^{(0)}=\f{\p}{\p
a^{(0)}}\mathcal {F}^{(0)}.\ee Near $u=1$, the electric theory is
strongly coupled. The electric-magnetic duality works in the sense
that if the theory is reformulated in terms of the dual magnetic
fields, it is weakly coupled. The dual prepotential
$\mathcal{F}_D^{(0)}$ can be obtained from \be a^{(0)}=\f{\p}{\p
a_D^{(0)}}\mathcal {F}_D^{(0)}.\ee A similar mechanism works for the
dyonic region near $u=-1$.

The Nekrasov theory can be viewed as the quantized version of the
Seiberg-Witten theory. The partition function
$Z(a,m,\epsilon_1,\epsilon_2,\Lambda)$ can be written in terms of
the prepotential $\mathcal{F}(a,m,\epsilon_1,\epsilon_2,\Lambda)$:
\be
Z(a,m,\epsilon_1,\epsilon_2,\Lambda)=\exp{\f{\mathcal{F}(a,m,\epsilon_1,\epsilon_2,\Lambda)}{\epsilon_1\epsilon_2}},\ee
where $a$ is related to the vacuum expectation value of scalar
fields, $m$ denotes the masses of matter fields, and
$\mathcal{F}(a,m,\epsilon_1,\epsilon_2,\Lambda)$ is a regular
function in the limit $\epsilon_1\to0, \epsilon_2\to0$. We are
interested in the case of $\epsilon_1=\hbar, \epsilon_2=0$ and with
no matter. It is shown in \cite{MM0910, WH1004} that the modul $a$
and the prepotential can be expanded as\beq
a(a^{(0)})&=a^{(0)}+\hbar^2a^{(1)}(a^{(0)})+\hbar^4a^{(2)}(a^{(0)})+\cdots\nn\\
\mathcal{F}(a,\hbar)&=\mathcal{F}^{(0)}(a)+\hbar^2\mathcal{F}^{(1)}(a)+\hbar^4\mathcal{F}^{(2)}(a)+\cdots,\eeq
where the superscript $(0)$ indicates the ``classical'' quantities
and the superscript $(i), i\ge1$ indicates the ``quantum'' corrected
ones. Note that the function variable of $\mathcal {F}$ is $a$
rather than $a^{(0)}$. A dual variable of $a$ can be defined by \be
a_D=\f{\p}{\p a}\mathcal {F}\ee and expanded as \be
a_D=a_D^{(0)}(a)+\hbar^2a_D^{(1)}(a)+\hbar^4a_D^{(2)}(a)+\cdots. \ee
In the limit $\hbar\to0$, only the leading order remains and it is
just the Seiberg-Witten theory. The higher order $\hbar$ corrections
are explained as effects of $\Omega$-twist in the gauge theory side,
and as quantization on the dual integrable system side. In
\cite{MM0910} the authors find $a^{(i)}(a^{(0)})$ and $a_D^{(i)}(a),
i\ge1$ can be obtained from $a^{(0)}(u)$ and $a_D^{(0)}(u)$ by
acting on certain higher order differential operators on them. In
the following, we will explain it and write the results in
hypergeometric function.

The integrals of (\ref{SWaad}) can be written in another form. If we
change the variable as $x=\cos\phi$, then the integrals become
$-(2\pi)^{-1}\int\sqrt{2(u-\cos\phi)}d\phi$, which is the classical
action integral $\int p(\phi)d\phi$ of the sine-Gordon action
$\mathcal {L}=\dot{\phi}^2-\cos\phi$ for a particle at the given
``energy'' $u$. In order to quantize the system, we are led to the
Schr\"{o}dinger equation \be
(-\f{\hbar^2}{2}\f{d^2}{d\phi^2}+\cos\phi)\Psi(\phi)=u\Psi(\phi),\label{eq:sch}\ee
When the system is quantized, the contour integrals (\ref{SWaad})
are lifted to the monodromies of the phase of the wave function
along the circles $\alpha$ and $\beta$. The Eq (\ref{eq:sch}) is the
Mathieu equation, and some properties of the corresponding gauge
theory have been obtained by analyzing its periodic solution
\cite{WH1004}. We apply WKB method and write the wave function as\be
\Psi(\phi)=\exp{\f{i}{\hbar}\int^{\phi}P(\phi^{'})d\phi^{'}}=\exp{\f{i}{\hbar}\int^{\phi}(P_0+\hbar
P_1+\hbar^2P_2+\cdots)d\phi^{'}},\ee then we have\beq
P_0&=&\sqrt{2(u-\cos\phi)},\qquad
P_1=\f{i}{2}(\ln P_0)^{'},\nn\\
P_2&=&-\f{1}{8P_0}[2(\ln P_0)^{''}-((\ln P_0)^{'})^2],\qquad
P_3=\f{i}{2}(\f{P_2}{P_0})^{'},\nn\\ \cdots\eeq where the prime
denotes $\f{\p}{\p\phi}$.

As $P_1$ and $P_3$ are total derivatives, their contour integrals
are zero. Only the contour integrals of $P_0,P_2,P_4,\cdots$ give
nonzero results, they are related to
$a^{(0)},a^{(1)},a^{(2)},\cdots$, respectively. The nonzero part of
the $P_2$ contour integral gives\beq \oint_{\alpha,\beta}
P_2d\phi&=&\f{1}{32\sqrt{2}}\oint_{\alpha,\beta}\f{\sin^2\phi-4u\cos\phi+4}{(u-\cos\phi)^{5/2}}d\phi\nn\\
&=&-\f{1}{48\sqrt{2}}\oint_{\alpha,\beta}\f{\cos\phi}{(u-\cos\phi)^{3/2}}d\phi\nn\\
&=&\f{1}{48}(2ud_u^2+d_u)\oint_{\alpha,\beta}\sqrt{2(u-\cos\phi)}d\phi,\label{2ndorder}\eeq
where $d_u$ denotes $\f{d}{du}$. Using the formula\be
\f{d}{dz}F(\alpha,\beta,\gamma;z)=\f{\alpha\beta}{\gamma}F(\alpha+1,\beta+1,\gamma+1;z),\ee
we get \beq
a^{(1)}&=&\f{1}{48}(2ud_u^2+d_u)a^{(0)}(u)\nn\\
&=&\f{1}{24}[(2(u+1))^{-3/2}F(-\f{1}{2},\f{1}{2},1;\f{2}{u+1})\nn\\
&\quad&-2(u-1)(2(u+1))^{-5/2}F(\f{1}{2},\f{3}{2},2;\f{2}{u+1})\nn\\
&\quad&-6u(2(u+1))^{-7/2}F(\f{3}{2},\f{5}{2},3;\f{2}{u+1})]\label{a1}\eeq
 \beq
a_D^{(1)}&=&\f{1}{48}(2ud_u^2+d_u)a_D^{(0)}(u)\nn\\
&=&\f{i}{96}[F(\f{1}{2},\f{1}{2},2;\f{1-u}{2})-\f{1}{16}(5u-1)F(\f{3}{2},\f{3}{2},3;\f{1-u}{2})\nn\\
&\quad&+\f{3}{16}u(u-1)F(\f{5}{2},\f{5}{2},4;\f{1-u}{2})].\label{aD1}\eeq

In a similar way, the third order contour integral is \be \oint
P_4d\phi=\f{1}{2^9\times 45}(28u^2d_u^4+120ud_u^3+75d_u^2)\oint
P_0d\phi.\label{4thorder}\ee We will not give the full detail here
because it is little long. Both $a^{(2)}$ and $a_D^{(2)}$ can be
written in a similar form as that of (\ref{a1}) and (\ref{aD1}) in
terms of hypergeometric functions. We can either expand them near
the point $u=\infty$ and $u=1$, or directly act on the forth order
differential operator on the series expansion of $a^{(0)}(u)$ and
$a_D^{(0)}(u)$, they will give the same result. This strategy can be
applied to higher order contour integrals, which contains more
complicated differential operators.

To obtain results (\ref{2ndorder}) and (\ref{4thorder}), we have
followed the trick of \cite{MM0910}, although our differential
operators are slightly different from theirs because we have set
$\Lambda=1$.

\section{Electric expansion}

In the following three sections we will derive the prepotential in
the electric region and the magnetic(dyonic) region. We will not
list the full details of the procedure, we only explicitly give some
series expansions which are interesting for our project. Some of
them have been known before, but appear in different literatures, or
derived through other methods. We use the newly discovered relation
between the gauge theory and the integrable model to give a complete
and consistent derivation. Some of our results, especially the
magnetic(dyonic) expansion of the prepotential for the case
$\epsilon_1=\hbar,\epsilon_2=0$, are new.

On the moduli space, $u=\infty$ corresponds to the electric region,
near this point the massless excitations are weakly coupled Abelian
U(1) electric fields. Expanding $a(u)$ and $a_D(u)$ near $\infty$,
up to the $\hbar^4$ order we have
\begin{align}
a(u)&=a^{(0)}(u)+\hbar^2a^{(1)}(u)+\hbar^4a^{(2)}(u)+\cdots\nn\\
&=\sqrt{2u}[1-\f{1}{4}(\f{1}{2u})^{2}-\f{15}{64}(\f{1}{2u})^{4}-\f{105}{256}(\f{1}{2u})^{6}-\f{15015}{16384}(\f{1}{2u})^{8}+\cdots]\nn\\
&\quad-\f{\hbar^2}{\sqrt{2u}}[\f{1}{2^4}(\f{1}{2u})^2+\f{35}{2^7}(\f{1}{2u})^4+\f{1155}{2^{10}}(\f{1}{2u})^6+\f{75075}{2^{14}}(\f{1}{2u})^8+\cdots]\nn\\
&\quad-\f{\hbar^4}{(2u)^{\f{3}{2}}}[\f{1}{2^6}(\f{1}{2u})^2+\f{273}{2^{10}}(\f{1}{2u})^4+\f{5005}{2^{11}}(\f{1}{2u})^6+\f{2297295}{2^8}(\f{1}{2u})^8+\cdots].\end{align}
This result has been obtained in \cite{MM0910} and \cite{WH1004},
through different methods. Our method here follows, and simplifies,
the one in \cite{MM0910}. From $a(u)$, the inverse series gives\beq
2u&=&a^2+\f{1}{2}a^{-2}+\f{5}{32}a^{-6}+\f{9}{64}a^{-10}+\f{1469}{8192}a^{-14}+\cdots\nn\\
&\quad&+\hbar^2(\f{1}{8}a^{-4}+\f{21}{64}a^{-8}+\f{55}{64}a^{-12}+\f{18445}{8192}a^{-16}+\cdots)\nn\\
&\quad&+\hbar^4(\f{1}{32}a^{-6}+\f{219}{512}a^{-10}+\f{1495}{512}a^{-14}+\f{985949}{65536}a^{-18}+\cdots).\label{typeAu}\eeq
The series expansion of
$a_D(u)=a_D^{(0)}+\hbar^2a_D^{(1)}+\hbar^4a_D^{(2)}$ is lengthy, it
contains many terms of the form $(c_1+c_2\ln2+c_3\ln
u)u^{k+\f{1}{2}}, k=0,1,2,\cdots$. Substituting $u=u(a)$ into
$a_D(u)$, we get the series expansion of $a_D(a)$ with very simple
structure. The prepotential is obtained from $a_D=\f{\p}{\p
a}\mathcal {F}$:
\beq \mathcal {F}(a,\hbar)&=&\f{i}{4\pi}[4a^2(\m{ln}2a-\f{3}{2})-\f{1}{2}a^{-2}-\f{5}{64}a^{-6}-\f{3}{64}a^{-10}+\cdots]\nn\\
&\quad&+\hbar^2\f{i}{4\pi}(\f{1}{6}\ln a-\f{1}{8}a^{-4}-\f{21}{128}a^{-8}-\f{55}{192}a^{-12}+\cdots)\nn\\
&\quad&+\hbar^4\f{i}{4\pi}(\f{1}{1440}a^{-2}-\f{1}{32}a^{-6}-\f{219}{1024}a^{-10}-\f{1495}{1536}a^{-14}+\cdots).\label{prepotE}\eeq

The results are consistent with other works. For example, in
\cite{MM0910}, the $\hbar^0$ and $\hbar^2$ order results of
$\mathcal {F}(a,\hbar)$ have been obtained through the same method
as here; in \cite{WH1004}, the form of power series of $\mathcal
{F}(a,\hbar)$ has been obtained through analyzing the Mathieu
function, here we derive the coefficients; in \cite{FP0208}, direct
gauge theory calculation gives the instanton part of the
prepotential up to four instantons contribution, it is easy to check
that our result is coincident with theirs if we set
$\epsilon_1=\hbar, \epsilon_2=0$. It is also worth mentioning that,
in \cite{FP0208}, other choices such as
$\epsilon_1=-\epsilon_2=\hbar$ or $\epsilon_1=\epsilon_2=\hbar$ will
give different results. Not only the rational coefficients are
different, the powers of $a$ are also different. For example, for
the case $\epsilon_1=-\epsilon_2$, the prepotential will be the one
given in \cite{KMT0206, HK0605} which is different from
(\ref{prepotE}). This fact explicitly indicates that the
quantization of the integrable model we discuss here indeed
corresponds to a special corner of the Nekrasov theory with
$\epsilon_1=\hbar, \epsilon_2=0$.

\section{Magnetic expansion}

Now we have confidence that the WKB contour integrals of the
integrable model indeed give the Nekrasov theory in the electric
region, what we will do next is just to expand the WKB contour
integrals in the magnetic region, i.e. near $u=1$ on the moduli
space. In the magnetic region, magnetic monopoles couple to the dual
U(1) Abelian gauge fields as massive matter hypermultiplets. The
effective action is obtained by integrating out all massive fields,
their effects are encoded in the subleading terms of (\ref{magFD}).

The motivation of studying magnetic expansion of Nekrasov function
comes from two sides.

First, the Nekrasov theory is formulated in the electric region,
where it can be explicitly expanded in terms of the electric
quantity $a$; however, its magnetic expansion is much less known,
although various dual quantities can be formally defined. A special
corner of the parameter space $\epsilon_1=-\epsilon_2$ has been
investigated in \cite{HK0605}, using results of the holomorphic
anomaly equation of topological string theory. However, the general
case is unknown. In this paper, through the relation with the
integrable system, we can explore another corner with
$\epsilon_1=\hbar, \epsilon_2=0$.

Second, the electric-magnetic duality of the gauge theory has some
interplay with duality of the integrable model. According to the
discussion in \cite{FGNR9906}, on the symplectic manifold $\mathcal
{M}$ related to the integrable system, the Liouville's theorem
states that the symplectic form $\omega$ has a normal form locally
written in terms of coordinates $(I, \varphi)$:\be \omega=dI\wedge
d\varphi,\ee where $\varphi$ is the coordinate variable, and $I$ is
the action variable in the sense that the Hamiltonian is a function
of only $I$: $H=h(I)$. For the classical integrable model which
corresponds to the $\mathcal {N}=2$ gauge theory, the symplectic
manifold is the tangent space of the gauge theory moduli space,
$\mathcal {M}=\mathcal {C}\times T$, where $\mathcal {C}$ is the
complex plane related to the vacuum expectation value of the adjoint
complex scalar field, and $T$ is the torus related to the complex
gauge coupling $\tau$. The symplectic form is\cite{DW9510} \be
\omega=da^{(0)}(u)\wedge \f{dx}{y(u,x)},\ee where $y=y(u,x)$ is the
Seiberg-Witten curve. The Hamiltonian of the integrable system is
identified with the beta function of the prepotential of the gauge
theory, and $a^{(0)}$ is the action variable.

The gauge theory has electric-magnetic duality which maps
$\tau\to-\f{1}{\tau}$ and $a^{(0)}\to a_D^{(0)}$, and we can
formulate the theory as either electric theory or magnetic theory.
Therefore, in the magnetic formulation, the symplectic structure
discussed above is reformulated in the dual quantities. Near
$u=\infty$, the gauge theory is a weakly coupled electric theory,
the appropriate action variable is $a^{(0)}(u)$. While near $u=1$
the gauge theory is a weakly coupled magnetic theory, the
appropriate action variable is $a_D^{(0)}(u)$. On the $u$-plane, we
have $da^{(0)}\wedge da_D^{(0)}=0$, therefore there exists a {\em
potential} that maps $a^{(0)}$ and $a_D^{(0)}$ to each other: \be
a_D^{(0)}=\f{\p}{\p a^{(0)}}\mathcal {F}^{(0)}\ee or \be
a^{(0)}=\f{\p}{\p a_D^{(0)}}\mathcal {F}_D^{(0)}.\ee Depending on
the electromagnetic frame we work in, we choose one of $\mathcal
{F}^{(0)}$ and $\mathcal {F}_D^{(0)}$ as the potential. We say this
integrable system manifests the {\em action-action duality}.
$\mathcal {F}^{(0)}$ and $\mathcal {F}_D^{(0)}$ are dual to each
other, they are the Seiberg-Witten {\em prepotential} of the gauge
theory in the electric and magnetic region, respectively. The
magnetic expansion of the Seiberg-Witten theory has been known
\cite{L9611}, and investigating the quantum version is a natural
next step.

If the classical electric-magnetic duality has a well defined
quantum correspondence, then there should exist a dual pair
$\mathcal {F}(a,\hbar)$ and $\mathcal {F}_D(a_D,\hbar)$. For the
special case of SU(2) pure Yang-Mills theory with $\epsilon_1=\hbar,
\epsilon_2=0$, the quantum correction can be expressed in terms of
hypergeometric function through WKB series and can be analytically
studied in the magnetic region. This provides a glimpse to the dual
phase of the integrable system.

Expanding (\ref{aaD0}) near the magnetic point $u=1$ (set
$\sigma=u-1$, therefore $d_u=d_\sigma$), and using the differential
operators of (\ref{2ndorder}) and (\ref{4thorder}), we get the
asymptotic form of $\hat{a}_D=ia_D$ and $a$ up to the order of
$\hbar^4$ \beq
\hat{a}_D(\sigma)&=&\hat{a}_D^{(0)}(\sigma)+\hbar^2\hat{a}_D^{(1)}(\sigma)+\hbar^4\hat{a}_D^{(2)}(\sigma)+\cdots\nn\\
&=&-\f{1}{2}\sigma+\f{1}{32}\sigma^2-\f{3}{512}\sigma^3+\f{25}{16384}\sigma^4+\cdots\nn\\
&\quad&-\f{\hbar^2}{2^7}(1-\f{5}{16}\sigma+\f{35}{256}\sigma^2-\f{525}{8192}\sigma^3+\cdots)\nn\\
&\quad&-\f{\hbar^4}{2^{18}}(-17+\f{721}{32}\sigma-\f{10941}{512}\sigma^2+\f{141757}{8192}\sigma^3+\cdots).\label{aDsigma}\eeq
the inverse series gives \beq
\sigma&=&-2\hat{a}_D+\f{1}{4}\hat{a}_D^2+\f{1}{32}\hat{a}_D^3+\f{5}{512}\hat{a}_D^4+\cdots\nn\\
&\quad&+\f{\hbar^2}{2^6}(-1-\f{3}{8}\hat{a}_D-\f{17}{64}\hat{a}_D^2-\f{205}{1024}\hat{a}_D^3+\cdots)\nn\\
&\quad&+\f{\hbar^4}{2^{17}}(9+\f{405}{16}\hat{a}_D+\f{2943}{64}\hat{a}_D^2+\f{69001}{1024}\hat{a}_D^3+\cdots).\label{typeBsigma}\eeq
In the magnetic region, the series expansion of $a(\sigma)$ contains
many terms of the form $(c_1+c_2\ln2+c_3\ln\sigma)\sigma^k,
k=0,1,2,\cdots$. Similar to the case of electric expansion, after
substituting $\sigma=\sigma(\hat{a}_D)$ into $a(\sigma)$, we get the
series expansion of $a(\hat{a}_D)$ with very simple structure. The
dual prepotential $\mathcal {F}_D$ can be obtained from $a=\f{\p}{\p
a_D}\mathcal {F}_D$: \beq \mathcal
{F}_D(a_D,\hbar)&=&\f{1}{i\pi}[\f{\hat{a}_D^2}{2}\ln(-\f{\hat{a}_D}{2})+4\hat{a}_D-\f{3}{4}\hat{a}_D^{2}
+\f{1}{16}\hat{a}_D^{3}+\f{5}{512}\hat{a}_D^{4}+\f{11}{4096}\hat{a}_D^{5}+\cdots]\nn\\
&\quad&+\f{\hbar^2}{i\pi2^5}(\f{1}{3}\ln\hat{a}_D-\f{3}{2^3}\hat{a}_D-\f{17}{2^7}\hat{a}_D^2
-\f{205}{2^{10}\times3}\hat{a}_D^3+\cdots)\nn\\
&\quad&+\f{\hbar^4}{i\pi2^{11}}(-\f{7}{45}\hat{a}_D^{-2}+\f{135}{2^9}\hat{a}_D+\f{2943}{2^{13}}\hat{a}_D^2+\cdots).\label{magFD}\eeq

Some interesting features of the dual prepotential (\ref{magFD}) can
be compared to that appearing in \cite{HK0605}, although they
discuss a different corner of the Nekrasov theory with
$\epsilon_1=-\epsilon_2$(see the Conclusion). Except the only two
terms containing $\ln\hat{a}_D$, the quantum corrections are powers
of $\hat{a}_D$, their coefficients are all rational numbers, the
same as that in \cite{HK0605}. This fact serves a nontrivial
examination of the result itself. If there were mistakes in the
coefficients of expansions $\hat{a}_D(\sigma),\sigma(\hat{a}_D)$ and
$a(\sigma)$, then (\ref{magFD}) would contain terms like
$c_1+c_2\ln2+c_3\ln\hat{a}_D$ in any other terms. In the $\hbar^4$
order correction of $\mathcal {F}_D$, the first two terms of order
$\mathcal {O}(\hat{a}_D^{-1})$ and $\mathcal {O}(1)$ are absent. The
same pattern happens in formula (2.33) of \cite{HK0605}. Actually,
in their case the ``gap'' phenomenon happens for all higher genus
corrections, we believe that in our case the ``gap'' phenomenon also
persists to higher order $\hbar$ corrections.

Although no direct gauge theory calculation in the dual magnetic
fields is available, however, from the experience of electric
expansion, we have an explanation for the different terms of
$\mathcal {F}_D$. Terms of order $\hat{a}_D^2\ln\hat{a}_D$ and
$\hbar^2\ln\hat{a}_D, \hbar^4\hat{a}_D^{-2},\cdots$ in the
prepotential correspond to the classical and one loop contributions.
Other terms correspond to integrating out multiparticle massive
hypermultiplets, which are monopole particles in the original
electric theory.

In \cite{NS0908, NS0901}, the prepotential of the gauge theory is
identified with the Yang-Yang function \cite{YY1969} of the quantum
Toda integrable model. The problem has two kinds of formulations,
called type A and type B spectral problems. The type B problem is
solved by the periodic Mathieu function with quantization
condition\cite{NS0908}\be \f{1}{\hbar}\f{\p}{\p
a_D}\mathcal{F}_D=\f{a}{\hbar}=n, \qquad n\in
\mathbb{Z}.\label{BetheB}\ee This is studied in \cite{WH1004}. The
type A problem is solved by the dual equation \be
\f{i}{\hbar}\f{\p}{\p a}\mathcal{F}=\f{\hat{a}_D}{\hbar}=m, \qquad
m\in \mathbb{Z}.\label{BehteA}\ee The two quantization conditions
serve as the Bethe equations of the corresponding spectrum problems.
However, the appearance of potential in (\ref{BetheB}) and
(\ref{BehteA}) only serves as a conceptual definition, in practice,
only $a(u)=\hbar n$ and $\hat{a}_D(\sigma)=\hbar m$ are used to
compute the energy spectrum $u$. The type B's eigenvalue as a
function of the quantum number $n$ is given in (\ref{typeAu}) as
series expansion, it can be expressed in a more compact form as the
periodic solution of the Mathieu equation \cite{WH1004}. The type
A's eigenvalue as a function of the quantum number $m$ is given in
(\ref{typeBsigma}), in the next section we will further explain its
relation to the Mathieu equation. The two type problems are
connected by the S duality $\tau\to-\f{1}{\tau}$.

Therefore we have a clear picture about the role of the
electric-magnetic duality on the side of the integrable model: it is
the action-action duality \cite{FGNR9906} of the quantum integrable
model that maps type A and type B spectrums to each
other\cite{NS0908}.

\section{Dyonic expansion}

The untwisted $\mathcal {N}=2$ SU(2) gauge theory has a global
$\mathbb{Z}_2$ symmetry acting on the $u$ plane by $u\to-u$. Under
the $\mathbb{Z}_2$ symmetry the magnetic region at $u=\Lambda^2$ is
mapped to the dyonic region at $u=-\Lambda^2$. At the dyonic point,
the massless soliton particles are either charge $(n_e,n_m)=(1,-1)$
dyons or charge $(n_e,n_m)=(1,1)$ dyons, depending on the direction
from which we cross the wall of marginal stability and approach the
dyonic point\cite{Ferrari9602}. We choose the convention
$(n_e,n_m)=(1,-1)$, therefore, the Seiberg-Witten solution behaves
as $a-a_D\sim u+1$ near $u=-1$. The electric-magnetic duality
together with the $\mathbb{Z}_2$ symmetry generate the
electric-magnetic-dyonic triality. In the following, we will give
the dyonic expansion of the Nekrasov theory, it is related to the
magnetic expansion by a $\mathbb{Z}_2$ symmetry of the Nekrasov
theory.

The dyonic expansion is very similar to that of the magnetic
expansion, therefore we only briefly report the main results.
Setting $\varpi=u+1$ we get \beq
a_T&=&a-a_D=a_T^{(0)}+\hbar^2a_T^{(1)}+\hbar^4a_T^{(2)}+\cdots\nn\\
&=&\f{1}{2}\varpi+\f{1}{32}\varpi^2+\f{3}{512}\varpi^3+\f{25}{16384}\varpi^4+\cdots\nn\\
&\quad&+\f{\hbar^2}{2^7}(1+\f{5}{16}\varpi+\f{35}{256}\varpi^2+\f{525}{8192}\varpi^3+\cdots)\nn\\
&\quad&+\f{\hbar^4}{2^{18}}(17+\f{721}{32}\varpi+\f{10941}{512}\varpi^2+\f{141757}{8192}\varpi^3+\cdots).\eeq
The inverse series gives the eigenvalue of the corresponding quantum
mechanics problem \beq
\varpi&=&2a_T-\f{1}{4}a_T^2-\f{1}{32}a_T^3-\f{5}{512}a_T^4+\cdots\nn\\
&\quad&+\f{\hbar^2}{2^6}(-1-\f{3}{2^3}a_T-\f{17}{2^6}a_T^2-\f{205}{2^{10}}a_T^3+\cdots)\nn\\
&\quad&+\f{\hbar^4}{2^{17}}(-9-\f{405}{2^4}a_T-\f{2943}{2^6}a_T^2-\f{69001}{2^{10}}a_T^3+\cdots).\label{typeBvarpi}\eeq
We can substitute $\varpi=\varpi(a_T)$ into either $a({\varpi})$ or
$a_D(\varpi)$, and the dyonic prepotential can be obtained from
either $a(a_T)=\f{\p}{\p a_T}\mathcal {F}_T$ or $a_D(a_T)=\f{\p}{\p
a_T}\mathcal {F}_T$. The two kinds of choice correspond to doing
electric-dyonic duality and magnetic-dyonic duality, respectively.
We choose electric-dyonic duality and get
\beq \mathcal{F}_T(a_T,\hbar)&=&\f{1}{i\pi}[\f{a_T^2}{2}\ln(-\f{a_T}{16})+4a_T-\f{3}{4}a_T^2+\f{1}{16}a_T^3+\f{5}{512}a_T^4+\f{11}{4096}a_T^5+\cdots]\nn\\
&\quad&+\f{\hbar^2}{i\pi2^5}(-\f{1}{3}\ln a_T+\f{3}{2^3}a_T+\f{17}{2^7}a_T^2+\f{205}{2^{10}\times3}a_T^3+\cdots)\nn\\
&\quad&+\f{\hbar^4}{i\pi2^{11}}(-\f{7}{45a_T^2}+\f{135}{2^{9}}a_T+\f{2943}{2^{13}}a_T^2+\cdots).\eeq
The dyonic prepotential is very close to that of magnetic, only
differing by a $-\f{3}{2}a_T^2\ln2$ term and a minus sign in the
$\hbar^2$ correction.

Now, we will discuss a relation between the magnetic(dyonic)
expansion and the periodic solution of the Mathieu equation.
Equation (\ref{eq:sch}) can be rewritten as\be
\Psi^{''}(z)+(A-2B\cos2z)\Psi(z)=0\ee with $A=\f{8u}{\hbar^2},
B=\f{4\Lambda^2}{\hbar^2}, 2z=\phi$, and $\Lambda$ restored. The
periodic solution is marked by a quantum number $\nu$, $\nu$ is an
even integer for the solution with period $\pi$, an odd integer for
the solution with period $2\pi$. In \cite{WH1004} we have studied
the eigenvalue problem of the periodic solution in the case of small
$\f{\sqrt{B}}{\nu}$ expansion, which corresponds to the electric
expansion of gauge theory. Here, we will encounter the small
$\f{\nu}{\sqrt{B}}$ expansion.

The eigenvalue formulas (\ref{typeBsigma})and (\ref{typeBvarpi})
have very similar structure, restore $\Lambda$ and set
$\hat{a}_D=\hbar\f{\nu}{2}$ in (\ref{typeBsigma}), and set
$a_T=\hbar\f{\nu}{2}$ in (\ref{typeBvarpi}), with $\nu$ an even
integer. Then (\ref{typeBsigma})and (\ref{typeBvarpi}) can be
rewritten in terms of $A, B, \nu$, and we find that they are
asymptotic expansions of the following two more compact expressions:
\beq
A_\nu&=&2B-4\nu\sqrt{B}+\f{4\nu^2-1}{2^3}+\f{4\nu^3-3\nu}{2^6\sqrt{B}}\nn\\
&\quad&+\f{80\nu^4-136\nu^2+9}{2^{12}B}+\f{528\nu^5-1640\nu^3+405\nu}{2^{16}B^{\f{3}{2}}}\nn\\
&\quad&+\f{2016\nu^6-10080\nu^4+5886\nu^2-243}{2^{19}B^2}+\cdots\label{eigenMag}\eeq
for (\ref{typeBsigma}), and \beq
A_\nu&=&-2B+4\nu\sqrt{B}-\f{4\nu^2+1}{2^3}-\f{4\nu^3+3\nu}{2^6\sqrt{B}}\nn\\
&\quad&-\f{80\nu^4+136\nu^2+9}{2^{12}B}-\f{528\nu^5+1640\nu^3+405\nu}{2^{16}B^{\f{3}{2}}}\nn\\
&\quad&-\f{2016\nu^6+10080\nu^4+5886\nu^2+243}{2^{19}B^2}+\cdots\label{eigenDy}\eeq
for (\ref{typeBvarpi}). Formulas (\ref{eigenMag}) and
(\ref{eigenDy}) can be found in the formula (20.2.30) in
\cite{AbramowitzStegun}. Their notation $w$ is related to ours by
$w=2\nu$. The terms of order $\f{1}{B^2}$ come from the $\hbar^6$
correction, we have checked all other higher order terms which we do
not explicitly list here for the sake of avoiding unnecessary
lengthy. The two cases are related to each other by the
$\mathbb{Z}_2$ symmetry $\nu\to i\nu, B\to -B$. To ensure the
expansions make sense, we need the ratio of the adjacent terms to be
small $\f{\nu}{\sqrt{B}}\sim\f{a_{D/T}}{\Lambda}<<1$, which is the
physical requirement that the expansions are performed in the weakly
coupled region of magnetic/dyonic theory.

The sine-Gordon model (\ref{eq:sch}) is the quantum mechanics
problem on the moduli space of the Seiberg-Witten theory. Its
excitation spectrum can be classified into three species: the
excitations in the bottom of the potential are ``dyonic
excitations''; the excitations near the mouth of the potential are
``magnetic excitations''; the excitations far above the top of the
potential are ``electric excitations''.

\section{Conclusion}

The recently discovered relation between the Nekrasov gauge theory,
i.e. $\mathcal {N}=2$ gauge theory in the $\Omega$ background, and
quantization of the algebraic integrable system is an exciting field
that needs deeper understanding. In this paper we study the
expansion of the SU(2) Nekrasov theory with $\epsilon_1=\hbar,
\epsilon_2=0$ on the whole moduli space, through its relation to the
sine-Gordon quantum mechanics model. We focus on this relatively
simple model because in this case many quantities can be explicitly
calculated, however, it presents the basic features of the novel
correspondence. The consistence of theses results on both sides,
i.e. on the gauge theory side and the integrable model side, gives
nontrivial support to the correspondence. Using the observation of
\cite{MM0910}, higher order quantum effects can be obtained by
acting on certain higher order differential operators on the
classical results, and can be compactly expressed in terms of the
hypergeometric function. The hypergeometric function is well defined
on the whole moduli space, therefore studying the property in the
magnetic (dyonic) region is straightforward, by expanding the
results near the magnetic(dyonic) points $u=\pm1$. It is remarkable
that the coefficients of the subleading terms of the prepotentials
$\mathcal {F}, \mathcal {F}_D$ and $\mathcal {F}_T$ are all rational
numbers. We stress here that the three prepotentials $\mathcal {F},
\mathcal {F}_D$ and $\mathcal {F}_T$ are the local asymptotic
expansions of the same object that is globally well defined. It
seems that some symmetries of the Seiberg-Witten theory, such as the
$\Gamma(2)$ modular symmetry and the $\mathbb{Z}_2$ global symmetry,
survive under the $\Omega$ twist with $\epsilon_1=\hbar,
\epsilon_2=0$.

The electric-magnetic duality of the gauge theory corresponds to the
action-action duality on the integrable system side. The
action-action duality exchanges the role of $a$ and $a_D$(or $a_T$)
and maps two kinds of spectrum problem to each other. The
prepotential $\mathcal {F}$ of the gauge theory serves as the
Yang-Yang function of the type A spectrum problem, and the dual
prepotential $\mathcal {F}_D$(and $\mathcal {F}_T$) serves as the
Yang-Yang function of the type B spectrum problem. For the case of
pure SU(2) Yang-Mills theory, the eigenvalue of the two types of
problems are well known results of the Mathieu equation.

In the four dimensional $\Omega$-deformed theory the two parameters
$\epsilon_1,\epsilon_2$ can take arbitrary complex value, the full
structure of the theory is very rich. Some special corners of the
parameter space have been studied in detail, they are often related
to some other field theory models. We briefly list several cases
that have appeared in the literature:
\\$\bullet$ The case of
$\epsilon_1=-\epsilon_2=\lambda$ has been studied in its related
contexts of topological string and matrix models, see for example
\cite{KMT0206}. In that case, the partition function takes the form
\be \mathcal {F}=\sum_{g=0}^{\infty}\lambda^{2g-2}\mathcal
{F}^{(g)}(a).\ee The higher order correction terms $\mathcal
{F}^{(g)}(a)$ correspond to the gravitational couplings to the
Seiberg-Witten gauge theory. $\mathcal {F}^{(g)}(a)$ is the
holomorphic limit of the quantity $F^{(g)}(\tau,\bar{\tau})$ of the
type B topological sigma model which corresponds to the higher genus
gravity correction. In \cite{HK0605} the authors study the SU(2)
theory, using the $\Gamma(2)\subset SL(2,\mathbb{Z})$ (quasi)modular
property of $F^{(g)}(\tau,\bar{\tau})$, the magnetic expansion of
$\mathcal {F}_D^{(g)}(a_D)$ was obtained by the limit
$\bar{\tau}_D=-\f{1}{\bar{\tau}}\to\infty$.
\\$\bullet$ The case of $\epsilon_1=\hbar, \epsilon_2=0$ is related to the quantization of integrable systems, initiated in \cite{NS0908}.
The present work investigates the particular SU(2) pure gauge theory
using its connection with the integrable system. Comparing our
results with that in \cite{HK0605}, we know that the Nekrasov theory
with $\epsilon_1=-\epsilon_2\ne0$ and $\epsilon_1\ne0,\epsilon_2=0$
will give the same $\mathcal {F}^{(0)}(a)$ and $\mathcal
{F}_D^{(0)}(a_D)$ which is the Seiberg-Witten theory, but for higher
order corrections, $\mathcal {F}^{(g)}(a)$ and $\mathcal
{F}_D^{(g)}(a_D)$ for $g\ge1$, they two cases give different
results.
\\$\bullet$ The case $\epsilon_1=\epsilon_2$ is
explored in \cite{pestun}. The corresponding gauge theory in the
$\Omega$-background is identical to the physical theory defined on
the Euclidean $S^4$, with $\epsilon_1=\epsilon_2$ equal to the
inverse of the radius of the four sphere.
\\$\bullet$ Recently, a relation between the four dimensional $\Omega$-deformed theory and
the two dimensional Liouville conformal field theory(CFT) is
discovered by Alday, Gaiotto and Tachikawa \cite{AGT}, it states
that the Nekrasov partition function is identical to the correlation
function of Liouville CFT on certain Riemann surface with punctures.
In their case, the corresponding Nekrasov theory sits in the corner
of $\epsilon_1\cdot\epsilon_2=1$, the $\epsilon_1,\epsilon_2$
parameters are related to the central charge of the Liouville CFT.
\\$\bullet$ More recently, in \cite{kreflWalcher}, the authors try to embed the deformed gauge theory with general $\epsilon_1, \epsilon_2$ into
topological string theory. Its partition function with generic
$\epsilon_1, \epsilon_2$ satisfies an extended version of the
holomorphic anomaly equation. Especially, they find that theory at
$\epsilon_1=-2\epsilon_2$ can be identified with an ¡°orientifold¡±
of the theory at $\epsilon_1=-\epsilon_2$.

It seems that the Nekrasov theory is a very powerful structure that
unifies several fields in an unexpected way. Exploring the general
case of $\epsilon_1, \epsilon_2$, especially its modular property,
largely remains untouched.

\section*{Acknowledgments}
W.H. would like to thank organizers and lecturers of a related
workshop held at ITP-CAS. Y-G.M was supported in part by the
National Natural Science Foundation of China under Grant
No.10675061. We thank the anonymous referee for pointing out results
of Refs \cite{pestun} and \cite{kreflWalcher} and their role in the
landscape of $\Omega$-deformed theories.

\end{document}